\documentstyle[preprint,aps,eqsecnum]{revtex}
\begin{document}
\draft
\preprint{\small Preprint CGPG-95/10-4  gr-qc/9602050}

\title{Einstein's equations and the chiral model}

\author{Viqar Husain}

\address{Department of Mathematics and Statistics,
University of Calgary\\
Calgary, Alberta T2N 4N1, Canada,\\
and\\
Center for Gravitational Physics and Geometry,
Department of Physics,\\
The Pennsylvania State University,\\
University Park, PA 16802-6300, USA.\footnote{Present address.
Email: husain@phys.psu.edu}}

\maketitle

\begin{abstract}

 The vacuum Einstein equations for spacetimes with two commuting
spacelike Killing field symmetries are studied using the Ashtekar
variables. The case of compact spacelike hypersurfaces which are
three-tori is considered, and the determinant of the Killing two-torus
metric is chosen as the time gauge. The Hamiltonian evolution
equations in this gauge may be rewritten as those of a modified SL(2)
principal chiral model with a time dependent `coupling constant', or
equivalently, with time dependent SL(2) structure constants.  The
evolution equations have a generalized zero-curvature formulation.
Using this form, the {\it explicit} time dependence of an infinite
number of spatial-diffeomorphism invariant phase space functionals is
extracted, and it is shown that these are observables in the sense
that they Poisson commute with the reduced Hamiltonian.  An infinite
set of observables that have SL(2) indices are also found.  This
determination of the explicit time dependence of an infinite set of
spatial-diffeomorphism invariant observables amounts to the solutions
of the Hamiltonian Einstein equations for these observables.

\end{abstract}
\bigskip
\pacs{PACS numbers: 04.20.Cv, 04.20.Fy, 04.60.Ds}
\vfill
\eject

\section{Introduction}

In classical field theory one would like to find physically
interesting solutions and to study the question of integrability.
This is a difficult problem for four-dimensional non-linear theories
such as Einstein gravity, and most of the integrable theories are two
dimensional. The classic examples are the Kortweg-de Vries and
Sine-Gordon models, whose study has led to systematic methods for
addressing the integrability question for two-dimensional field
theories \cite{das,fadtakh}.  Self-dual Yang-Mills and gravity
theories are the only ones in four dimensions which are considered to
be integrable. This is because of twistor constructions of the general
solutions \cite{pen,ward1}. However, self-dual theories are unusual in
that they have formulations as two dimensional theories, and so the
standard two dimensional methods are likely to be applicable.  For
example, self-dual gravity has a formulation as the two-dimensional
principal chiral model \cite{vsd1,vsd2,ward2}.

Vacuum Einstein gravity with two commuting spacelike Killing vector
fields is a two-dimensional field theory which has two local degrees
of freedom. Perhaps the main result in this reduction of Einstein
gravity is the discovery by Geroch of an infinite dimensional `hidden
symmetry' of the field equations \cite{geroch}. This symmetry leads to
a solution generating method for this sector of the Einstein
equations, which has since been studied from various points of view
\cite{kinnchit,hausern,wu,belzak,nenadb}.

As for other two-dimensional theories, hidden symmetries suggest that
this reduction of the Einstein equations is integrable.  However there
is so far no proof of integrability in the Liouville sense - that is,
it has not been shown that there exists an infinite number of Poisson
commuting constants of the motion. In order to address this question,
one would first have to identify the hidden symmetry generators on the
phase space, or otherwise give a solution to the Hamiltonian equations
of motion.

Apart from the purely classical reasons, classical integrability is
very useful for contructing quantum theories. This is because in
canonical quantization, one would like to convert gauge invariant
phase space functionals into Hermitian operators on a suitable
representation space. In generally covariant theories it is difficult
to find a sufficiently large number of such functionals. If the
classical theory is known to have hidden symmetries, then it should be
possible to find them explicitly on the phase space. The phase space
generators of the symmetries would then serve as observables.

In this paper we study the two commuting spacelike Killing field
reduced equations from a Hamiltonian point of view. This will be with
a view to addressing the integrability question and identifying
explicitly the hidden symmetries on the phase space.  We will use the
Ashtekar canonical formalism \cite{ash,abhbook} for complex
relativity, where the conjugate phase space variables are complex, but
are defined on a real manifold.

There has been previous work by the author on this reduction using the
Ashtekar variables \cite{v1}, where it was claimed that the
Hamiltonian evolution equations were equivalent to those of the
SL(2) principal chiral model. It was then pointed out \cite{mizo}
that this identification was based on a further reduction of the
theory, because the gauge fixing used in it could not be achieved for
generic two Killing field reduced spacetimes. In particular it could
not be achieved for the Gowdy models \cite{gow}. In this paper we will
show, among other things, how to rectify this situation by using a
gauge fixing that was suggested in Ref. \cite{v1}.

The outline of this paper is as follows: In the next section we review
the reduction of the Einstein theory to the sector where the metric
has two commuting spacelike Killing fields. Details of this may be
found in a previous work by the author and Smolin \cite{lv}.  In
Section III we focus on the three-torus Gowdy cosmology, and fix the
determinant of the Killing two-torus metric as the time gauge. In this
gauge the Einstein evolution equations become those of a {\it
modified} (complexified) SL(2) principal chiral model {\it with a time
dependent `coupling constant'}.  Section IV shows how the evolution
equations may be written as a {\it generalized} zero-curvature
equation. The generalization is in the fact that our equation has
explicit time dependence whereas the standard zero curvature equations
for integrable two-dimensional models do not. Section V describes how
to extract the time dependence of a specific infinite set of
diffeomorphism invariant observables on the phase space, and identify
explicitly the generators of the hidden symmetries of the reduced
Hamiltonian.  The last section gives the main conclusions, and
contains a discussion of the relevance of the results for constructing
a quantum theory for this sector of Einstein gravity.

\section{Two Killing vector field reduction}

 The Ashtekar Hamiltonian variables \cite{ash,abhbook} for
complexified general relativity
 are the (complex) canonically conjugate pair $(A_a^i, \tilde{E}^{ai})$
where $A_a^i$ is an SO(3) connection and $\tilde{E}^{ai}$
is a densitized dreibein. $a,b,..$ are three dimensional spatial indices,
 $i,j,..=1,2,3$ are internal SO(3) indices, and the tilde denotes a
density of weight one. The constraints of general relativity are
\begin{eqnarray}
{\cal G}^i &:=& D_a\tilde{E}^{ai} = 0,  \\
{\cal C}_a &:=& F_{ab}^i\tilde{E}^{ai}=0,  \\
{\cal H} &:=& \epsilon^{ijk}F_{ab}^i\tilde{E}^{aj}\tilde{E}^{bk}=0,
\end{eqnarray}
where
\begin{equation}
D_a\lambda^i = \partial_a \lambda^i + \epsilon^{ijk}A_a^j\lambda^k
\end{equation}
is the covariant derivative, and
\begin{equation}
F_{ab}^i = \partial_a A_b^i -
\partial_b A_a^i + \epsilon^{ijk} A_a^j A_b^k
\end{equation}
is its curvature.

Since the phase space variables are complex, reality conditions need
to be imposed to obtain the Euclidean or Lorentzian sectors. These are
\begin{equation}
A_a^i=\bar{A}_a^i \ \ \ \ \  E^{ai}=\bar{E}^{ai}
\label{euc}
\end{equation}
 for the former, and
\begin{equation}
A_a^i + \bar{A}_a^i = 2\Gamma_a^i(E),\ \ \ \ \   E^{ai}=\bar{E}^{ai}
\label{lor}
\end{equation}
 for the latter. The $\Gamma_a^i(E)$ is the connection for spatial
 indices and the bar denotes complex conjugation.

 We now review the two commuting spacelike Killing field reduction of
these constraints which was first presented in \cite{lv}. Working in
spatial coordinates $x,y$, such that the Killing vector fields are
$(\partial/\partial x)^a$ and $(\partial/\partial y)^a$ implies that
the phase space variables will depend on only one of the three spatial
coordinates. Specifically,  we assume that the spatial topology is that
of a three torus so that the phase space variables depend on the time
coordinate  $t$ and one angular coordinate $\theta$. This situation
corresponds to one of the Gowdy cosmological models \cite{gow}.
 (The other permitted spatial topologies for the Gowdy cosmologies
 are $S^1\times S^2$ and  $S^3$.)

In addition to these Killing field conditions, we set to zero some
of the phase space variables as a part of the symmetry reduction:
\begin{eqnarray}
\tilde{E}^{x3}&=&\tilde{E}^{y3}=\tilde{E}^{\theta 1}=
\tilde{E}^{\theta 2}=0,
 \nonumber \\
 A_x^3 &=& A_y^3=A_\theta^1 = A_\theta^2 = 0.
\end{eqnarray}
These conditions may be viewed as implementing a partial gauge fixing
and solution of some of the resulting second class constraints.
Details of these steps are given in Ref. \cite{lv}. The end result,
Eqns. (\ref{G}-\ref{H}) below,
is a simplified set of first class constraints that describe a two
dimensional field theory with two local degrees of
freedom on $S^1\times R$.
Renaming the remaining variables
$A:=A_\theta^3$, $E:=\tilde{E}^{\theta 3}$ and  $A_\alpha^I$,
$\tilde{E}^{\alpha I}$, where $\alpha,\beta,..=x,y$ and $I,J,..=1,2$,
the reduced constraints are
\begin{eqnarray}
 G & := &\partial E + J =0, \label{G} \\
 C  & := & A\partial E - \tilde{E}^{\alpha I}
\partial A_\alpha^I = 0, \label{C} \\
  H & := &-2\epsilon^{IJ}F_{\theta\alpha}^I\tilde{E}^{\alpha J}E
+ F_{\alpha\beta}\tilde{E}^{\alpha I} \tilde{E}^{\beta J}
\epsilon_{IJ} \nonumber \\
  & = &-2 E \tilde{E}^{\alpha J} \epsilon^{IJ}\partial A^I_\alpha
 +2AEK - K_\alpha^\beta K_\beta^\alpha + K^2 = 0 \label{H},
\end{eqnarray}
where $\partial=(\partial/\partial\theta)=\prime$,
\begin{eqnarray}
K_\alpha^\beta&:=&A_\alpha^I \tilde{E}^{\beta I}, \ \ \ \
K:=K_\alpha^\alpha, \\
J_\alpha^\beta&:=&\epsilon^{IJ} A_\alpha^I \tilde{E}^{\beta J},
\ \ \ \ J:=J_\alpha^\alpha,
\end{eqnarray}
and $\epsilon^{12}=1=-\epsilon^{21}$.

The SO(3) Gauss law has been reduced to U(1), and the spatial
diffeomorphism constraint to Diff($S^1$). This  may be seen by
calculating the Poisson algebra of the  constraints smeared by
functions $\Lambda(t,\theta)$, the shift $V(t,\theta)$, and the
lapse $N(t,\theta)$ (which is a density of weight $-1$). With
\begin{eqnarray}
G(\Lambda) &=& \int_0^{2\pi} d\theta\ \Lambda G\ , \\
C(V) &=& \int_0^{2\pi} d\theta\ VC\ , \\
H(N) &=&\int_0^{2\pi} d\theta\ NH \ ,
 \end{eqnarray}
the constraint algebra is
\begin{eqnarray}
\{G(\Lambda),G(\Lambda')\} &=& \{G(\Lambda),H(N)\}=0,\\
\{C(V),C(V')\}&=&C({\cal L}_VV' ), \\
\{H(N),H(N')\}&=& C(W) - G(AW) ,
\end{eqnarray}
where
\begin{equation}
W\equiv E^2(N\partial N' - N'\partial N).  \label{stfn}
\end{equation}
This shows that $C$ generates Diff$(S^1)$. Also we note that this
 reduced first class
system still describes a sector of general relativity due to
the Poisson bracket $\{H(N),H(N')\}$,  which is the reduced version
of that for full general relativity  in the Ashtekar variables.

The variables $K_\alpha^\beta$ and $J_\alpha^\beta$ defined above
will be used below in the discussion of observables. Here we note their
properties. They are invariant under the reduced Gauss law (\ref{G}),
transform as densities of weight +1 under the Diff($S^1$) generated
by $C$, and form the   Poisson algebra
\begin{eqnarray}
\{K_\alpha^\beta,K_\gamma^\sigma\} &=&
\delta_\alpha^\sigma K_\gamma^\beta
 - \delta_\gamma^\beta K_\alpha^\sigma, \\
\{J_\alpha^\beta,J_\gamma^\sigma\} &=&
-\delta_\alpha^\sigma K_\gamma^\beta
 + \delta_\gamma^\beta K_\alpha^\sigma, \\
 \{K_\alpha^\beta,J_\gamma^\sigma\} &=&
\delta_\alpha^\sigma J_\gamma^\beta
 - \delta_\gamma^\beta J_\alpha^\sigma.
\end{eqnarray}
This shows that  $K_\alpha^\beta$ form the gl(2) Lie algebra, and hence
generate gl(2) rotations on variables with indices $\alpha,\beta,..=x,y$.

The following linear combinations of $K_\alpha^\beta$ form the sl(2)
subalgebra of gl(2):
 \begin{equation} L_1 = {1\over 2}(K_y^x + K_x^y), \ \ \
  L_2 = {1\over 2}(K_x^x - K_y^y), \ \ \
  L_3 = {1\over 2}(K_y^x - K_x^y) \ .
\label{sl2r}
 \end{equation}
The Poisson bracket algebra of these is
\begin{equation} \{L_i,L_j\} = C_{ij}^{\ \ k}L_k \ ,
 \end{equation}
 where $C_{12}^{\ \ 3} = -1, C_{23}^{\ \ 1} = 1, C_{31}^{\ \ 2} = 1$
are the sl(2) structure constants. (From here on the indices $i,j,k...$
will denote sl(2) indices, and not the so(3) internal indices of the
Ashtekar formulation).
The corresponding linear combinations  of $J_\alpha^\beta$ are denoted
 by $J_i$, $i=1,2,3$. Their Poisson brackets  are
 \begin{equation}
 \{L_i,J_j\}=C_{ij}^{\ \ k}J_k,\ \ \
 \{J_i,J_j\}= - C_{ij}^{\ \ k}L_k.
 \end{equation}
 We also have
 \begin{equation}
 \{J,J_i\}=\{J,L_i\}=\{K,J_i\}=\{K,L_i\}= 0.
 \end{equation}
For discussing observables, it will turn out to be very convenient to
replace the eight canonical phase space variables
$A_\alpha^I$ and $\tilde{E}^{\alpha I}$ by the eight Gauss law
invariant variables $K_\alpha^\beta$ and $J_\alpha^\beta$. We will
refer to the latter as the gl(2) variables.

\section{Gauge fixing}

In this section we pick a preferred foliation of the two-Killing
field reduced spacetimes by fixing the time coordinate. We will show,
by giving the spacetime metric, that the  time gauge choice is
  the one made by Gowdy\cite{gow}. It will turn out that the
evolution equations in this gauge become those of a modified
 SL(2) chiral model which has a time dependent `coupling constant'.

Since the theory we are considering is complex general relativity on
a real manifold, we need to fix, as the (real) time coordinate,
a (complex) phase space variable that transforms as a scalar under
the spatial diffeomorphisms (\ref{C}). Since $E$ transforms like
 a scalar in the reduction we are considering,  we set
\begin{equation}
{\rm Im}(E)=0, \ \ \ \   {\rm Re}(E) = t
\end{equation}
as the time gauge choice. The (complex) reduced Hamiltonian density
is by definition the negative of the variable conjugate to time
($ = E$):
 \begin{equation}
H_R:= -A = -{1\over K} \tilde{E}^{\alpha J} \epsilon^{IJ}\partial
A^I_\alpha
+ {1\over 2Kt}\ \bigl( K^2 - K_\alpha^\beta K_\beta^\alpha \bigr).
\label{RH}
\end{equation}
The lapse density  $ N$  is determined by requiring that the
gauge fixing condition be preserved in time:
 \begin{equation}
\dot{E} = \dot{t} = 1 = \{ E, H(N)\},
\end{equation}
 gives
\begin{equation}
 N = -{1 \over 2 t K}.\label{laps}
\end{equation}
With this gauge choice, we also find that the Gauss law
constraint (\ref{G}) reduces further to
\begin{equation}
J=0,
\end{equation}
and remains first class.

The evolution equations of the gl(2) variables in this gauge
are derived from Hamilton's equation
\begin{equation}
\dot{X} = \{X, \int_0^{2\pi} d\theta\ [\ H_R  + \Lambda\ J\ ]
+ C(V) \},
\label{fullH}
\end{equation}
where  $C(V)$  generates spatial diffeomorphisms and  $J$ generates
the Gauss law rotations.
 In the following we focus only on the contributions from $H_R$,
because below we will make a $\theta$ coordinate fixing that sets the
shift $V$ to zero. Also, the Gauss law term in the full Hamiltonian
(\ref{fullH}) does not contribute to the evolution of the gl(2)
variables, because these variables are already Gauss law invariant.

The equations for $J$ and $K$ are
\begin{eqnarray}
\dot{J} &=& 0, \\
\dot{K} &=& \partial({J\over K})=0,
 \label{evo4}
\end{eqnarray}
where the last equality follows because of the reduced Gauss law
in this gauge. These imply that the Gauss law is preserved under
evolution, and that $K=f(\theta)$, where $f$
is an arbitrary density on the circle.

The equations for $L_i$ and $J_i$ are
\begin{eqnarray}
\dot{L}_i &=& \partial ({J_i \over K}), \label{evo1} \\
\dot{J}_i &=& -\partial ({ L_i\over K})
   -  {2 \over t K}\ C_i^{\ jk}J_j L_k. \label{evo2}
\end{eqnarray}
The explicit time dependence of the reduced Hamiltonian appears
only in the $\dot{J}_i$ equation.

These evolution equations may be further simplified by
setting a  $\theta$ coordinate condition which  makes
the spatial diffeomorphism constraint (\ref{C}) second class.
As for the time gauge fixing above, a scalar phase space function
must be fixed as the $\theta$ coordinate.
 Since $K$ transforms as a density of weight one, we can set
\begin{equation}
 \theta = -{1\over \alpha}\ {\rm Re}(\int^\theta d\theta'\ K(\theta',t) ,
\ \ \ \  {\rm Im}(\int^\theta d\theta'\ K(\theta',t)=0.
\end{equation}
This gives $K=-\alpha\ne 0$, a constant density. We note that a scalar
density transforms non-trivially as $K'(\theta',t)=
\partial\theta/\partial\theta'\ K(\theta,t)$ under coordinate
transformations.  Therefore, setting a density on the spatial surface
to be a constant indeed fixes a spatial coordinate.\footnote{ This
type of coordinate fixing condition - setting a phase space density on
the circle to be a constant, has been used in the quantization of the
one polarization Gowdy cosmology by Berger \cite{bb}.} We also note
that the constant of motion $\int_0^{2\pi} d\theta\ K(\theta,t)$ now
takes the value $-2\pi\alpha$.

The shift vector $V$ is determined by requiring that the gauge
condition be preserved under Hamiltonian evolution. We have
\begin{equation}
\dot{\theta} = 0 = \{ -{1\over \alpha} \int^\theta d\theta'\
K(\theta',t)\ ,\ \int_0^{2\pi}d\theta\ [\ H_R + \Lambda\ J\ ] + C(V) \}
= -{1\over \alpha} \int^\theta d\theta'\ {\cal L}_V  K(\theta',t),
\end{equation}
where ${\cal L}_V$ denotes the Lie derivative, and the last
equality follows because
$\{\ \int_0^{2\pi}d\theta\ H_R,\ J\ \}=
\{\ \int_0^{2\pi}d\theta\ H_R,\ K\ \}=0$. Therefore $V=0$.

The evolution equations (\ref{evo1}-\ref{evo2}) for $L_i$ and $J_i$,
which are the only non-trivial ones, then become\footnote{
We note that $\int_0^{2\pi}d\theta\ H_R$ is a first class
Hamiltonian because it commutes with the second class pair consisting of
the diffeomorphism constraint and the $\theta$ coordinate fixing
condition. Therefore the Dirac brackets give the same
evolution equations as the ordinary Poisson brackets.}
\begin{eqnarray}
\dot{L}_i + {1\over \alpha} J_i'&=& 0  \\
\dot{J}_i -{1\over \alpha} L_i' + {2 \over \alpha t }\
C_i^{\ jk}L_j J_k&=& 0.
\end{eqnarray}
We now make a further change of variable, the time rescaling
$\tau = t/\alpha$, and then replace $\tau$ by $t$. This changes the
above equations to
\begin{eqnarray}
\dot{L}_i +  J_i'&=& 0 \label{con} \\
\dot{J}_i - L_i' + {2 \over \alpha t }\ C_i^{\ jk}L_j J_k&=& 0. \label{cur}
\end{eqnarray}
These resemble the first order form of the evolution equations of
the principal chiral model, which for the two-dimensional Lie algebra
valued gauge field $A_\mu$ ($\mu=x,t$) are
\begin{eqnarray}
  \partial_\mu A_\mu & = &  0, \nonumber \\
   \partial_\mu A_\nu - \partial_\nu A_\mu +
[ A_\mu , A_\nu ]  & = &  0. \label{chiral}
\end{eqnarray}
The difference from the latter is the $1/t$ factor in (\ref{cur}),
which is like a time dependent coupling constant. This factor may be
absorbed in the structure constants to make them time dependent. Thus,
this reduction of the Einstein equations is equivalent to an SL(2)
chiral model in which the structure constants scale like $1/t$.

Of the original eight gl(2) variables we have the six sl(2)
variables $J_i$ and $L_i$ left.  The two conditions still to be
imposed to gauge fix completely are the spatial diffeomorphism
constraint
\begin{equation}
 \tilde{E}^{\alpha I}\partial A_\alpha^I = 0,
\end{equation}
which is now second class, and a gauge condition for
fixing the remaining Gauss law, $J=0$. These will reduce
the six sl(2) variables down to the four phase space degrees
of freedom per point for gravity. We will
not reduce the system completely by solving these two
conditions,  but rather focus on studying  the
symmetries  associated with the evolution equations
(\ref{con}-\ref{cur}).

We now give the spacetime metric that results from the time
gauge fixing described above. The doubly densitized inverse of the
spatial metric   in the Ashtekar formulation is
${\tilde{\tilde{q}}}^{ab}=\tilde{E}^{ai} \tilde{E}^{bi}$.
 Therefore, with $q\equiv {\rm det}(q_{ab})$, the spatial metric
is given by
\begin{equation}
{\tilde{\tilde{q}}}^{ab} = q q^{ab} =
\left( \begin{array}{cc}
  E^2 & 0 \\
  0  & e^{\alpha\beta} \end{array}
\right).
\end{equation}
where $e^{\alpha\beta}:=\tilde{E}^{\alpha I} \tilde{E}^{\beta I}$.
The determinant of the spatial metric is
$ q = t \sqrt{{\rm det}(e^{\alpha\beta})}=:t e $.
Therefore
\begin{equation}
 q_{ab} =  \left( \begin{array}{cc}
 e/t & 0 \\
  0  & t e e_{\alpha\beta} \end{array}
\right).
\end{equation}
>From (\ref{laps}), the lapse function ${\cal N}$ is
\begin{equation}
{\cal N} = \sqrt{q} N =  \sqrt{{e\over 4t}}.
\end{equation}
 The line element (after the above rescaling $t\rightarrow t\alpha$)
 in terms of phase space variables is then
\begin{equation}
ds^2 =  {e(t,\theta)\over \alpha t}\ (\ -{ 1\over 4}\ dt^2  +
d\theta^2\ )    + \alpha t\ e(t,\theta)\ e_{\alpha\beta}(t,\theta)
\ dx^\alpha dx^\beta.
\label{metric}
\end{equation}
>From this we see that the determinant of the Killing two-torus metric
is det$(\alpha t e e_{\alpha\beta})=(\alpha t)^2$.  Since
$e_{\alpha\beta}$ is complex the metric is also. The real Lorentzian
section of this is determined by the reality condition
Im$(e_{\alpha\beta})=0$. This gives the Gowdy $T^3$ metric. We note
that $t=0$ is the initial spacelike cosmological singularity, and that
the time gauge choice $E=t$ turns out to be the same as Gowdy's gauge
\cite{gow}, namely the determinant of the Killing two-torus metric is
time.

\section{Evolution equations as a `zero-curvature' condition}

The zero curvature formulation for a non-linear field theory, which is
basically the same as the equation for the Lax pair for the theory,
arises from a linear system of equations which are also known as the
inverse scattering equations. The latter is a pair of equations whose
integrability condition gives the non-linear field theory in
question. This formulation is important for determining integrability
\cite{das,fadtakh} because the conservation laws associated with zero
curvature equations are relatively easy to obtain.  All known
integrable models have such formulations.  Our form below
(\ref{zcurv}) for the two Killing field reduced Einstein equations is
different from all the other known models in that it contains explicit
time dependence.  This is a direct consequence of the fact that we
have the time dependent reduced Hamiltonian (\ref{RH}). Nevertheless,
as we will see in the following section, an infinite set of symmetries
of the reduced Hamiltonian can still be obtained as a consequence of
(\ref{zcurv}).

The evolution equations (\ref{con}-\ref{cur}) derived in the last
section may be rewritten in a compact form using the sl(2) matrix
generators
\begin{equation}
   g_1 = {1\over 2\alpha }\left( \begin{array}{cc}
         0 & 1 \\
         1 & 0 \end{array} \right)\ , \ \ \
   g_2 = {1\over 2\alpha }\left( \begin{array}{cc}
         1 & 0 \\
         0 & -1 \end{array} \right)\ ,\ \ \
   g_3 = {1\over 2\alpha }\left( \begin{array}{cc}
         0 & 1 \\
	 -1 & 0 \end{array} \right)\ ,
\end{equation}
which satisfy  the relations
\begin{equation}
[\ g_i,\ g_j\ ]= {1\over \alpha}\  C_{ij}^{\ \ k}g_k\ ,\ \ \ \ \
g_i\ g_j={ 1\over 2\alpha }\ C_{ij}^{\ \ k}g_k\ .
\end{equation}
Defining the matrices
\begin{equation}
 A_0 := 2 L_ig_i\ , \ \ \ \ \ \ A_1 := 2 J_ig_i\ ,
 \end{equation}
  the evolution equations (\ref{con}-\ref{cur}) become
\begin{equation}
\partial_0 A_0 + \partial_1 A_1 = 0
 \label{cons}
 \end{equation}
\begin{equation}
\partial_0 A_1 - \partial_1 A_0 + {1\over t}\ [\ A_0,\ A_1\ ] = 0.
\label{curv}
\end{equation}
Equations (\ref{cons}-\ref{curv}) resemble the first order form of the
SL(2) chiral model field equations, but with the `coupling constant'
factor $1/t$ multiplying the commutator.

The two evolution equations (\ref{cons}-\ref{curv}) may be rewritten
as a single equation in the following way. Define for a
parameter $\lambda$
\begin{equation}
a_0:= {1\over 1+\lambda^2}\ \bigl( A_0 - \lambda A_1 \bigr) \ \ \
a_1:= {1\over 1+\lambda^2}\ \bigl( \lambda A_0 +  A_1 \bigr)
\label{ai}
\end{equation}
Then equations (\ref{cons}-\ref{curv}) follow from the
single time dependent `zero-curvature' equation
\begin{equation}
\dot{a}_1 - a_0' + {1\over t}\ [\ a_0,\ a_1\ ] = 0.
 \label{zcurv}
\end{equation}
 This equation is the main result of this section, and represents a
compact way of writing the evolution equations for this reduction
of general relativity.\footnote{We note that in the derivation of
these equations we used the fundamental Poisson bracket relation for
Euclidean complexified general relativity $\{A_a^i(x),E^b_j(y)\} =
\delta_a^b \delta^i_j \delta(x,y)$, without the factor $i$ of the
Lorentzian Ashtekar variables on the right hand side. When this factor
is put in one gets the Lorentzian chiral model equation, where a relative
minus sign appears in (\ref{con}) above. All the results of this paper
go through with appropriate  sign changes in the definitions (\ref{ai}).}

\section{Observables}

In this section, the dynamical equation (\ref{zcurv}) is used to
obtain the observables, which are the phase space functionals that
Poisson commute with the reduced Hamiltonian (\ref{RH}).  We note that
the standard procedure that applies to two-dimensional models
\cite{das,fadtakh} now has to be modified since our zero curvature
equation (\ref{zcurv}) has explicit time dependence. We also note that
the standard (time independent) zero-curvature equations allow one to
establish two results: (1) The extraction of an infinite number of
phase space functionals that commute with the Hamiltonian, and (2) a
simple proof that these functionals are in involution. Below we
establish the first result for our system using the generalized
`zero-curvature' equation.

The transfer matrix used in the study of two-dimensional models is
like the Wilson loop in non-Abelian Yang-Mills theory.  The zero
curvature formulation of the evolution equations of a theory implies
that the trace of the transfer matrix is a conserved quantity, or
`observable'. This is true for essentially the same reason as that
which makes the trace of the Wilson loop an observable in 2+1 gravity
\cite{witt,2+1}.

We consider the following time dependent analog of the transfer
matrix
\begin{eqnarray}
U[A_0,A_1](0,\theta) &:=&
{\rm P exp}\bigl[\ {1\over t}\int_0^{\theta} d\theta\
a_1(t,\theta,\lambda)\ \bigr] \nonumber \\
&\equiv & I + {1\over t}\int_0^{\theta}\ d\theta'\ a_1(\theta,t,\lambda)
\nonumber \\
& & + {1\over t^2}\int_0^{\theta}\ d\theta' \int_0^{\theta'}
d\theta''\  a_1(\theta'',t,\lambda) a_1(\theta',t,\lambda) + ...,
\label{tran}
\end{eqnarray}
where $I$ is the $2\times 2$ identity matrix.
$U(0,\theta)$ depends on time explicitly, and also implicitly through
the gravitational variables. Defining
\begin{equation}
M={\rm Tr}\ U(0,2\pi) \label{tranM},
\end{equation}
we have
\begin{equation}
{dM\over dt} =  {\partial M\over \partial t} +
\int_0^{2\pi}\ d\theta\
{\delta M\over \delta a_{1j}(\theta,t)}
{\partial a_{1j}(\theta,t)\over\partial t}, \label{Mdot}
\end{equation}
where  $a_1 := a_{1i}g_i$.
The second term on the right hand side  is zero because
\begin{eqnarray}
\int_0^{2\pi}d\theta\ {\delta M\over \delta a_{1j}(\theta,t)}
{\partial a_{1j}(\theta,t)\over\partial t}&=&
\int_0^{2\pi} d\theta\ {\rm Tr}\ [\ U(0,\theta) {1\over t}
{\partial a_1\over \partial t} U(\theta,2\pi) \ ] \nonumber \\
&=& \int_0^{2\pi} d\theta\ {\rm Tr}\ [\ U(0,\theta) {1\over t}
({\partial a_0\over \partial \theta} -  {1\over t} [a_0,a_1])
 U(\theta,2\pi) \ ]\nonumber  \\
&=&
{1\over t^2}\int_0^{2\pi} d\theta\ {\rm Tr}\ [\ U(0,\theta)
(-a_1 a_0 + a_0 a_1 - [a_0,a_1])  U(\theta,2\pi)  \ ] \nonumber \\
& & +\ {1\over t}{\rm Tr}\ [\ U(0,2\pi)a_0(2\pi) - a_0(0)U(0,2\pi)\ ]
\nonumber \\
&=& 0 \label{impt},
\end{eqnarray}
where we have used (\ref{zcurv}), integrated by parts, and
used
\begin{equation}
U'(0,\theta) = {1\over t}\ U(0,\theta) a_1(\theta)\ \ \ \ \ \ \
 U'(\theta, 2\pi) = -{1\over t}\ a_1(\theta) U(\theta, 2\pi).
\end{equation}
The second term in the third equality in (\ref{impt}) is the surface
term arising from the integration by parts, which without the trace on
$M$ gives a non-zero contribution.

The time dependence of $M$ is given by Hamilton's equation
\begin{equation}
{d M\over dt} = \{ M,\  \int_0^{2\pi}d\theta\ H_R \} +
{\partial M\over \partial t},
\end{equation}
where the first term on the right hand side gives the implicit time
dependence. Therefore, the calculation (\ref{impt}) above gives
the crucial result
  \begin{equation}
  \{ M, \int_0^{2\pi}d\theta\ H_R\  \} = 0.
\label{M,H}
\end{equation}

Therefore each coefficient of $\lambda$ in $M$ is a phase space
functional that generates a symmetry of the reduced Hamiltonian.  This
is one of the main results of this paper. We note two points about the
observables generated using $M$ : (1) The observables, while being
symmetries of the reduced Hamiltonian, are not constants of the motion
because of their explicit time dependence, (indeed the reduced
Hamiltonian itself is explicitly time dependent), and (2) if the
Poisson bracket in (\ref{M,H}) is replaced by the Dirac bracket the
result is the same, because (as also noted above) the reduced
Hamiltonian is first class and $M$ is spatial-diffeomorphism
invariant. Furthermore, $M$ also has vanishing Poisson and Dirac
brackets with the remaining first class constraint $J$, again because
$M$ is spatial-diffeomorphism invariant, and because $J$ commutes with
$K$ (which is used in the $\theta$ fixing condition), and also
with the $J_i$ and $L_i$ (out of which $M$ is made). Therefore the
functionals generated via $M$ are indeed time dependent observables of
the theory.

There are  a set of three constants of the motion for this
system that have been given before \cite{lv}. These are
\begin{equation}
 l_i := \int_0^{2\pi} d\theta\  L_i(t,\theta).
\end{equation}
It is obvious from (\ref{con}) that these are conserved.

These  $l_i$ may be used to obtain an  infinite number of phase space
functionals with sl(2) indices that commute with the reduced
Hamiltonian $H_R$. These arise from the `generating functional'
\begin{equation}
\alpha_i := \{ l_i, M \}. \label{alpha}
\end{equation}
Expanding $\alpha_i$ in a power series in $\lambda$ gives
the infinite set of functionals $\alpha^n_i$ as coefficients of
$\lambda^n$. Explicitly
\begin{equation}
\alpha_i^n := {\partial^n\over \partial \lambda^n}
\{ l_i, M \}|_{\lambda = 0} .
\end{equation}
The first three of these are
\begin{eqnarray}
\alpha_i^0 &=& \{\ l_i, {\rm TrPexp}\int_0^{2\pi} d\theta\
A_1(\theta)\ \},\\
\alpha_i^1 &=& \{\ l_i,  {1\over t}\int_0^{2\pi} d\theta\
{\rm Tr}\ [\ V(0,\theta) A_0(\theta) V(\theta, 2\pi)\ ] \},\\
\alpha_i^2 &=& \{\ l_i, {1\over t}\int_0^{2\pi} d\theta\
{\rm Tr}\ [\ V(0,\theta) A_1 V(\theta, 2\pi)\ ] \nonumber \\
& & + {2\over t^2}\int_0^{2\pi} d\theta\int_0^{2\pi} d\theta'\
{\rm Tr}\ [\ V(0,\theta')A_0(\theta')V(\theta',\theta)A_0(\theta)
V(\theta,2\pi)\ ]\ \},
\end{eqnarray}
where
\begin{equation}
V(0,\theta):= {\rm TrPexp}\bigl[\ {1\over t}\int_0^\theta d\theta'\
A_1(\theta',t)\ \bigr].
\end{equation}
The functionals on the right hand sides in the Poisson brackets
resemble the loop variables in 3+1 gravity in the Ashtekar formulation
\cite{lc}. In the two Killing field reduction here, there is
effectively only the loop that wraps around the $\theta$
circle. However, unlike the 3+1 gravity loop variables, these Poisson
commute with the reduced Hamiltonian. We note that there is a factor
of $1/t$ associated with each insertion of $A_0$ or $A_1$ on the loop,
and that $n$ counts the number of such insertions.  This suggests an
affine algebra structure for the $\alpha_i^n$ Poisson algebra.

 What we have given are the observables in the Ashtekar variables for
the spacetime metric (\ref{metric}), which is the standard form for
metrics with two commuting spacelike Killing vector fields.  To
calculate the observables explicitly from a given spacetime metric of
the class we are considering is a straightforward procedure. The steps
are: (i) Calculate the extrinsic curvature $k_a^i$ and Christoffel
connection $\Gamma_a^i(E)$, which gives the Ashtekar connection $A_a^i
= \Gamma^a_i + i k_a^i$, (ii) calculate the sl(2) variables $L_i$
and $J_i$ (\ref{sl2r}), which gives the variable $a_1$ (\ref{ai}), and
finally, (iii) calculate the generating functional $M[a_i]$
(\ref{tranM}), whose expansion in powers of $\lambda$ gives all the
observables. This will give the observables explicitly as functions
of the spacetime metric variables, rather than as functions of
Hamiltonian variables.

\section{Reality conditions}

So far the theory we have been discussing is a reduction of complex
general relativity. Therefore the phase space observables given in the
last section are also complex. In order to obtain the observables for
the real Euclidean and Lorenztian theories we must impose the reality
conditions (\ref{euc}) and (\ref{lor}).

The Euclidean conditions simply imply that the phase space variables
must be real from the start. Hence the specialization of the
observables to this case is easy - we set $J_i$ and $L_i$ in the
generating functional $M$ (\ref{tranM}) to be real, which leads
directly to real observables.

For the Lorentzian theory, we note first that a real observable can
always be defined for the complex theory. To see this we first set
$E^{ai}$ to be real, which is one of the reality conditions.
The reality condition on $A_a^i$ implies that $\delta/\delta A =
-\delta/\delta \bar{A}$. This implies that the complex conjugate of an
observable is also an observable because the complex conjugates of the
constraints are also constraints.  Then if $O[E,A]$ is an
observable for the complex theory, a real observable for the
complex theory is $O[A,E] + O[\bar{A},E]$ \cite{thomas}. Therefore
the observables in the Lorentzian theory are given by
\begin{equation}
O_L[A,E] := (O[A,E] + O[\bar{A},E])|_{\bar{A} = 2\Gamma - A}\ .
\end{equation}

For our case, we can define the Lorentzian generating functional $M_L$
for the Lorentzian observables from $M$ (\ref{tranM}) in exactly this
way.  Equivalently, this can be done seperately for each observable
derived from $M$. Therefore the symmetries described above go through
for the Lorentzian theory as well.

\section{Conclusions and discussion}

There are three main results presented in this paper. The first is a
rewriting of the vacuum Einstein equations for metrics with two
commuting spacelike Killing vector fields, such that they resemble the
field equations of the SL(2) principal chiral model.  The only
difference from the latter is that the `coupling constant' is
explicitly time dependent. The second result is a further rewriting of
the reduced equations which leads to a generalized zero curvature
formulation. The third is the explicit identification of an infinite
set of phase space functionals which generate the hidden symmetries of
the reduced Hamiltonian via Poisson brackets. These phase space
functionals are spatial-diffeomorphism invariant and their time
dependence is  explicit. {\it This amounts to a solution of the
equations of motion for this infinite set of variables.}

We have not addressed the question of Liouville integrability for this
system, though the above results may provide a first step in this
direction.  The infinite set of phase space functionals given above
that commute with the Hamiltonian do not have non-vanishing Poisson
brackets with one another.  It has been shown in Ref. \cite{wu} that
the Lie algebra of the Geroch group is in fact the sl(2) affine
algebra. It is possible that the algebra of our observables is exactly
this, since they appear to form an sl(2) loop algebra.

For integrability one would like to show whether there are sums of
products of observables which do commute with one another. For the
two-dimensional models with standard (time independent) zero-curvature
equations, it is possible to show that there are two distinct
symplectic structures on their phase spaces. This fact leads to a
relatively easy proof of integrability \cite{das}.  In the present
case, with the understanding achieved so far, we do not know how to do
this.

Finding observables in the classical theory is a prerequisite for
certain quantization schemes. There is some debate concerning how
observables should be defined \cite{karel} in a generally covariant
theory. One view is that observables should be {\it fully} gauge
invariant, which means that they are constants of the motion or
`perennials'. The quantum theory would then be constructed by finding
suitable representations of the Poisson algebra of these observables.
This raises the question of how one would see time evolution in the
quantum theory, since constants of the motion do not evolve.  Another
view is that only {\it kinematically} gauge invariant functionals
should be used for quantization, and that the Hamiltonian constraint
should be converted into a functional Schrodinger equation.  In this
approach, if it can be carried through, time evolution would be seen
in the same way as for non-generally covariant theories.

The observables we have given fall into neither category because we
have fixed a specific time gauge.  The observables commute with the
reduced Hamiltonian, but also have explicit time dependence. Thus,
this situation appears to have the virtues of both the above
viewpoints. In particular, as stated above, these observables are
solutions to the equations of motion. A drawback may be that, although
two-volume may be a physically reasonable definition of time in
cosmology, the quantum theory would be dependent on this prefered
choice.

To proceed with quantization one would first need to know what the
algebra of the $\alpha_i^n$ is. Since each $n$ counts the number of
insertations of $A_0$ or $A_1$ in the $\theta$ circle, the algebra
structure already resembles that of an affine (Kac-Moody)
algebra. This is because the Poisson bracket of elements with $m$ and
$n$ insertions would lead to one with $m+n$ insertions.  We
conjecture, from the above similarity with the chiral model and the
result of Ref. \cite{wu}, that the algebra of certain sums of these
observables is the sl(2) affine algebra. The quantum theory would
then arise as a representation of this algebra. Since the time
evolution of each $\alpha_i^n$ is already known, the result would be
an evolving quantized algebra of observables. To see that this indeed
comes about is a topic for further work.

This paper has been restricted to the case of spacetimes that have
compact spatial surfaces. For the non-compact case one would have to
keep track of the boundary terms that arise in the constraints. In
particular there will be a surface contribution to the Hamiltonian.
Here also it would be of interest to find a gauge fixing that leads to
some generalized zero-curvature form of the evolution equations that
allows the extraction of physical observables.

\bigskip
I would like to thank Abhay Ashtekar, John Friedman and Lee Smolin for
discussions. This work was supported by the Natural Sciences and Engineering
Research Council of Canada, and by NSF grant PHY-93-96246 to the
Pennslyvania State University.


\begin{references}

\bibitem{das} A. Das, {\it  Integrable models}, (World Scientific,
 Singapore, 1989).

\bibitem{fadtakh} L. D. Faddeev and L. A. Takhtajan,
 {\it Hamiltonian methods in the theory of solitons},
 (Springer-Verlag, Berlin, 1987).

\bibitem{pen} R. Penrose, Gen. Rel. and Grav. {\bf 7}, 31  (1976).

\bibitem{ward1} R. S. Ward, Phys. Lett. {\bf A61}, 81, (1977);
M. Atiyah and R. S. Ward, Commun. Math. Phys. {\bf 55}, 117 (1977).

\bibitem{vsd1} V. Husain, Phys. Rev. Lett. {\bf 72}, 800 (1994);
V. Husain, Class. and Quantum Grav. {\bf 11}, 927 (1994).

\bibitem{vsd2} V. Husain, J. Math. Phys., to be published (1995),
hep-th/9410072.

\bibitem{ward2} R. S. Ward, Class. Quantum Grav. {\bf 7}, L217 (1990).

\bibitem{geroch} R. Geroch, J. Math. Phys. {\bf 13}, 394 (1971).

\bibitem{kinnchit} W. Kinnersley, J. Math. Phys. {\bf 18}, 1529 (1977);
W. Kinnersley and D. Chitre, J. Math. Phys. {\bf 18}, 1538 (1977).

\bibitem{hausern} I. Hauser and F. J. Ernst, Phys. Rev. D {\bf 20},
362, 1738 (1979); J. Math. Phys. {\bf 21}, 1126, 1418 (1980).

\bibitem{wu} Y. S. Wu and M. L. Ge, J. Math. Phys. {\bf 24}, 1187
(1983).

\bibitem{belzak} V. A. Belinskii and V. E. Zakharov, Sov. Phys.
JETP {\bf 48}, 985 (1978); {\bf 50}, 1 (1979).

\bibitem{nenadb} N. Manojlovic and B. Spence, `Integrals of the
motion intwo Killing field reduction of general relativity',
Syracuse University preprint SU-GP-93/7-8 (1993).

\bibitem{ash} A. Ashtekar, Phys. Rev. Lett. {\bf 57}, 2244 (1986);
Phys. Rev. D {\bf 36}, 1587 (1987).

\bibitem{abhbook}A. Ashtekar, {\it Non-perturbative canonical
gravity}, (World Scientific, Singapore, 1991).

\bibitem{v1} V. Husain, Phys. Rev. D {\bf 50}, 6207 (1994).

\bibitem{mizo} S. Mizoguchi,  Phys. Rev. D {\bf 51}, 6788 (1995).

\bibitem{gow} R. H. Gowdy, Ann. Phys. {\bf 83}, 203 (1974).

\bibitem{lv} V. Husain and L. Smolin, Nucl. Phys. {\bf B327}, 205
(1989).

\bibitem{bb} B. Berger, Ann. Phys. (NY) {\bf 83}, 458 (1974);
University of Maryland Ph.D. Thesis (1973).

\bibitem{witt} E. Witten, Nucl. Phys. {\bf B311 }, 46 (1988).

\bibitem{2+1} A. Ashtekar, V. Husain, C. Rovelli, J. Samuel and
L. Smolin, Class. Quantum Grav. {\bf 6}, L185 (1989).

\bibitem{lc} C. Rovelli and L. Smolin, Phys. Rev. Lett. {\bf 61},
1155 (1988); Nucl. Phys. {\bf B331}, 80 (1990).

\bibitem{thomas} Real observables for Lorentzian spherically symmetric
gravity in the Ashtekar variables have been discussed by
T. Thiemann and H. A Kastrup, Penn. State preprint (1994).

\bibitem{karel} K. V. Kuchar, in {\it Proceedings of the 13th
InternationalConference on General Relativity and Gravitation},
ed. C. Kozameh (IOP Publishing, Bristol, 1993).


\end{references}
\end{document}